\documentclass[amsmath,amssymb,aps,prd,showpacs,nofootinbib]{revtex4}
\usepackage{graphicx}
\usepackage{amsfonts}
\usepackage{amssymb}
\usepackage{amsmath}
\usepackage{bm}
\usepackage{latexsym}

\newcommand{\beq}{\begin{eqnarray}}
\newcommand{\eeq}{\end{eqnarray}}
\newcommand{\be}{\begin{equation}}
\newcommand{\ee}{\end{equation}}

\def\fun#1#2{\lower3.6pt\vbox{\baselineskip0pt\lineskip.9pt
\ialign{$\mathsurround=0pt#1\hfil ##\hfil$\crcr#2\crcr\sim\crcr}}}

\newcommand{{\SD}}{\rm SD}

\newcommand{\vep}{\bm p}

\begin{document}

\title{The radial Regge trajectories and leptonic widths of the
isovector mesons}

\author{\firstname{A.~M.}~\surname{Badalian}}
\email{badalian@itep.ru} \affiliation{Institute of Theoretical and Experimental
Physics, Moscow, Russia}

\author{\firstname{B.~L.~G.}~\surname{Bakker}}
\email{b.l.g.bakker@vu.nl} \affiliation{Department of Physics and Astronomy,
Vrije Universiteit, Amsterdam, The Netherlands}

\date{\today}

\begin{abstract}
It is shown that two physical phenomena are important for high
excitations: (i) the screening of the universal gluon-exchange
potential and (ii) the flattening of the confining potential owing
to creation of quark loops, and both effects are determined
quantitatively. Taking the first effect into account, we predict
the masses of the ground states with $l=0,1,2$ in agreement with
experiment. The flattening effect ensures the observed linear
behaviour of the radial Regge trajectories $M^2(n)=m_0^2 + n_r
\mu^2$ GeV$^2$, where the slope $\mu^2$ is very sensitive to the
parameter $\gamma$, which determines the weakening of the string
tension $\sigma(r)$ at large distances. For the $\rho$-trajectory
the linear behaviour starts with $n_r=1$ and the values
$\mu^2=1.40(2)$~GeV$^2$ for $\gamma=0.40$ and $\mu^2=1.34(1)$~GeV$^2$
for $\gamma=0.45$ are obtained. For the excited states the leptonic
widths: $\Gamma_{\rm ee}(\rho(775))=7.0(3)$~keV,  $\Gamma_{\rm
ee}(\rho(1450))=1.7(1)$~keV, $\Gamma_{\rm ee}(\rho(1900))=1.0(1)$~keV,
$\Gamma_{\rm ee}(\rho(2150))=0.7(1)$~keV, and $\Gamma_{\rm
ee}(1\,{}^3D_1)=0.26(5)$~keV are calculated, if these states are
considered as purely $q\bar q$ states. The width $\Gamma_{\rm
ee}(\rho(1700))$ increases if $\rho(1700)$ is mixed with the
$2\,{}^3S_1$ state, giving for a mixing angle $\theta=21^\circ$
almost equal widths: $\Gamma_{\rm ee}(\rho(1700))=0.75(6)$~keV and
$\Gamma_{\rm ee}(1450)=1.0(1)$~keV.  
\end{abstract}

\maketitle

\section{Introduction}
\label{sect.01}

Meson spectroscopy continues to be an important issue both for
experimentalists and theoreticians.  More precise experimental data
have appeared in the last years \cite{1,2,3,4,5,6,7,8,9} and a large
number of theoretical works are devoted to light-meson properties
\cite{10,11,12,13,14,15,16,17,18,19,20}. The important idea that
mesons have universal properties, from light-light to heavy quarkonia,
is supported in many studies \cite{21,22,23,24,25,26,27,28} and a
kind of universal  $q\bar q$ potential was used for all mesons in
different models \cite{9,10}, \cite{12,13,14,15,16}, \cite{22,23,24}.
The detailed analysis of meson spectra was done in the relativized
potential model (RPM), introducing a phenomenological (universal)
potential \cite{9}. A convenient systematics of radial excitations
was suggested in Ref.~\cite{11}, where it was assumed that the slope
of the radial Regge trajectories (RTs) has a universal value (with
a good accuracy) for all mesons. However, up to now the discussions
continue about the true value of the slopes of the radial RTs
\cite{24,25,26,27}, and even the linearity of the radial RTs is
disputed \cite{28}. However, the physical effects which are responsible
for the observed universality remain unclear up to now.

Here we use the relativistic string Hamiltonian (RSH) \cite{12,13},
derived in the framework of the field correlator method \cite{29,30},
which allows for expressing the meson properties via two fundamental
parameters: the string tension and the QCD constant $\Lambda$. In
principle, the RSH contains both perturbative and nonperturbative
dynamics, and yields also the spin-dependent interactions, so that,
in general, all possible dynamical regimes in the $q\bar q$ systems
can be addressed. It is the main purpose of our  work to start a
general analysis of the light meson dynamics both in radial and
orbital excitations. However, in the present paper we confine
ourselves to the case of the radial excitations of the vector mesons
$\rho(nS)$ and the ground states with $l=0,1,2$, where the physical
picture is more simple and transparent. Thus, our analysis can be
considered as the first step towards the overall picture, which may
be more complicated.

We show that in light mesons the dynamics is more complicated than
in heavy quarkonia, which manifests itself in two effects: the
so-called screening of the gluon-exchange (GE) interaction  and the
flattening of the linear confining potential, which is especially
important for high excitations. These phenomena occur for extended
objects owing to $q\bar q$ holes (loops), which are created inside
the film subtended by the Wilson loop. These two effects can be
described by the RSH and will be the main subject of our analysis.
As was shown in Ref.~\cite{12}, the RSH defines two regimes: the
string regime, valid for the states with large $l$, $l\geq 3$, and
the potential-like regime, taking place for low-lying states. For
the ground states (with large $l$)  the mass formula:  $M^2(l,n_r=0)
= 2\pi\sigma\sqrt{l(l+1)}$ was derived, which explicitly shows that
the slope of the leading RT is equal to $2\pi\sigma=(1.13\pm
0.02)$~GeV$^2$  with great accuracy (e.g. for $l=3$ the accuracy
is $0.8\%$). To derive this expression it was assumed that in RSH
the centrifugal term (the rotation of the string) gives a large
contribution, while the corrections to the mass from the GE, the
spin-dependent potentials, and the self-energy term are considered
to be small and may be neglected.  However, for low-lying states
these terms in the $q\bar q$ potential are not small \cite{22,23}
and therefore the question arises how to match the mass relations,
valid for the states with large $l$, and those with $l=0,1,2$.
First, we discuss the well-established features of the universal
$q\bar q$ interaction.

In the RSH approach the $q\bar q$  potential is defined in a
gauge-invariant way via the Wilson loop \cite{29,30} and the confining
potential is shown to be scalar and linear (if no quark loops occur):
$V_{\rm C}(r)=\sigma_0 r$, with the string tension
$\sigma_0=0.18(2)$~GeV$^2$ fixed by the slope of the leading RT
(with $j=l+s$) \cite{31,32}. In the static potential, the confining
and GE potentials enter as a sum to satisfy the Casimir scaling,
observed on the lattice with very good accuracy \cite{33,34}. A
very important point is that the parameters of the GE potential
cannot be taken arbitrarily but have to be determined in full
correspondence with the existing information from pQCD \cite{35}.
From high energy experiments the QCD constant
$\Lambda_{\overline{MS}}(n_f=5)$ is now well established, while the
QCD constants for $n_f=3,4$ are defined by matching the coupling
at the quark mass thresholds \cite{35,36,37}; it gives the value
$\Lambda_{\overline{MS}}(n_f=3)=(339\pm 10)$~MeV, if
$\alpha_s(M_Z)=0.1184(7)$ is used  \cite{35}, or a bit smaller
$\Lambda_{\overline{MS}}(n_f=3)=(327\pm 12)$~MeV is obtained for
the new world-average $\alpha_s(M_Z)=0.1177(13)$ \cite{37}. Knowledge
of $\Lambda_{\overline{MS}}(n_f=3)$ is very important, because its
value determines the ``vector" constant $\Lambda_V(n_f=3)$, entering
the vector coupling in the GE potential: $\Lambda_{\rm
V}(n_f=3)=1.4753~\Lambda_{\overline{MS}}(n_f=3)= (485\pm 25)$~MeV
\cite{36}. Besides, as shown recently in Ref.~\cite {38}, the
infrared regulator (IR) $M_{\rm B}$ is not an extra parameter, but
can be expressed via the string tension:
$M_B^2=2\pi\sigma=1.13(11)$~GeV$^2$ (the accuracy of calculations
is estimated to be $\sim 10\%$). Then taking the central values,
$\Lambda_V=0.485$~GeV and $M_B=1.13$~GeV, one defines the two-loop
freezing coupling (called {\em critical}), $\alpha_{\rm crit}=\alpha_{\rm
V}(q^2=0)=0.6065$.  Surprisingly, this value of the two-loop
$\alpha_{\rm crit}$ with a large $\Lambda_{\rm V}\sim 480$~MeV
coincides  with the one-loop phenomenological $\alpha_{\rm crit}$
from Ref.~\cite{9}, where a very small (unrealistic)
$\Lambda(n_f=3)=200$~MeV is used. Knowledge of this freezing constant
$\alpha_{\rm crit} = 0.60\pm 0.04$ is crucially important for heavy
quarkonia, where the GE interaction remains important up to high
excitations. Unfortunately, the role of the GE interaction in light
mesons is not fully understood and here we pay special attention
to the correct definition of the universal potential to distinguish
between true dynamical effects and artifacts coming from  different
fitting parameters, including the constituent masses.  Many features
of the light-meson dynamics become more transparent if one studies
the $S-$wave isovector mesons, which are more simple from the
theoretical point of view, since they are not subject to chiral
effects (with exception of the $\pi$-meson) and for them there does
not exist a complicate centrifugal term in the RSH.

We now pay special attention to the radial RTs with the systematics,
suggested in Ref.~\cite{11}, assuming that the radial RTs are linear in
the ($n_r,M^2$)-plane ($J^{PC}$ is fixed),
\begin{equation}
 M^2(n_r) = M^2(n_r=0) + n_r\mu^2.
\label{eq.1}
\end{equation}
Here  $M(0)$ is the mass of the lowest-lying meson on the RT and
$\mu^2$ is the slope parameter. According to Refs.~\cite{8,11} the
slope $\mu^2$ is approximately the same for all radial RT trajectories,
$\mu^2=1.25\pm 0.15$~GeV$^2$. This strong statement cannot be checked
in many cases, since no sufficient experimental information is
available about high radial excitations, with exception of the
$\rho$ family, and in the literature there are also other predictions
for $\mu^2$  \cite{18,25}, and even the linear behavior is disputed
\cite{28}. The question is whether this slope is universal or not.
Notice, that in different RPMs \cite{9,16,23} a much larger mass
difference $\mu_1^2= M^2(\rho(1450)) - M^2(\rho(775)) = (1.54\pm
0.04)$~GeV$^2$ is obtained and this value agrees with experimental
$\mu_1^2({\rm exp.})$, if the central values of the mass, $M(\rho(775))$
and $M(\rho'(1450))$ from the Partical Data Group (PDG) \cite{1}
are taken. An even larger value $\mu_1^2({\rm exp.)}=1.63(2)$~GeV$^2$
corresponds to the recent BaBar data for $M(\rho')=(1493\pm 15)$~MeV
\cite{5}. In the present paper we will show that this large mass
difference is not accidental and occurs because the $\rho(775)$
mass is ``too small'' due to a large GE and self-energy contributions.
For that reason (in contrast to other RTs) the linear behavior of
the radial $\rho$-trajectory starts with the first excitation
$n_r=1$.

Our calculations here are done in closed-channel approximation,
neglecting the widths and hadronic shifts, while in a strict sense
light mesons have to be studied as many-channel systems, taking
into account a contribution of every channel to the meson wave
function (w.f.). But such many-channel calculations form a very
difficult task, which needs individual consideration of every meson
and a complete theory of meson decays, which does not exist now.
Therefore, calculations in closed-channel approximation continue
to be very important: they allow for a separation of the conventional
$q\bar q$ mesons from multi-quark systems of a different nature
\cite{7}. Moreover, the influence of open channels can be effectively
taken into account, introducing the string tension $\sigma(r)$
depending on the separation $r$ \cite{23}. This effect occurs owing
to the creation of virtual quark loops in the Wilson loop, causing
the string tension to decrease and depend on $r$, and this effect
is very important for higher radial excitations,  while the ground
states are not affected by this {\em flattening} effect, since they
have relatively small sizes.

An important point is that one can introduce the critical value
of the string tension, $\sigma_{\rm crit}$, when the breaking of
the $q\bar q$ string takes place. If the string tension is taken
as in Ref.~\cite{23}: $\sigma(r) = \sigma_0 (1 - \gamma f(r))$ (with
$\sigma_0=0.18(2)$~GeV$^2$), then at not too large distances, $r\leq
1.2$~fm the string tension is almost constant, $\sigma(r)\approx
\sigma_0$, while at larger distances the function $f(r)\rightarrow
1$ and the critical value is
\begin{equation}
 \sigma_{\rm lim} = \sigma_0 (1 - \gamma).
\label{eq.2}
\end{equation}
The calculations show that a good description of the radial excitations
is reached if the parameter $\gamma=0.43\pm 0.03$ is used as a
fitting parameter. Moreover, the value of $\gamma$ strongly affects
the slope of the radial RT and therefore it can be extracted from
this slope, if there are good experimental data for the mass of the
excitations with $n_r\geq 2$. In particular, more precise data on
the masses of $\rho(1900)$ and $\rho(2150)$ could allow for
distinguishing between the value $\mu^2=1.43(13)$~GeV$^2$, suggested
in Ref.~\cite{18}, and $\mu^2=1.365(108)$ obtained in Refs.~\cite{8,11}
from the analysis of the Crystal Barrel data \cite{2}.

Here we also calculate the leptonic widths of the $\rho(n\,{}^3S_1)$
and $\rho(n\,{}^3D_1)$ states. However, the accuracy of these calculations is
limited by the fact that they are done in closed-channel approximation,
where the norm of the $q\bar q$ component of the w.f. at the origin
remains undetermined, e.g. for the states with $J^{PC}=1^{--}$ the
w.f. can be schematically written as
\begin{eqnarray}
 \psi_{\rm S}(r) & = & C_{q\bar q}(\cos\theta~ \psi_{\rm S}(r)
 -  \sin\theta~ \psi_{\rm D}(r)) + C_{\rm cont}(S) \psi_{\rm CS},
\nonumber \\
 \psi_{\rm D}(r) & = & C_{q\bar q}(\cos\theta~ \psi_{\rm D}(r)+
 \sin\theta ~\psi_{\rm S}(r))  + C_{\rm cont}(D) \psi_{\rm CD},
\label{eq.3}
\end{eqnarray}
assuming that the $q\bar q$ components of the $S$- and $D$-wave
w.f.s have equal (or close) values and allowing for $S-D$ mixing.
Fortunately, knowledge of the continuum component is not important
for the leptonic widths, since a multi-quark component of the w.f.,
even if it is large, gives a small contribution to the w.f. at the
origin \cite{39}. Thus the weight $C_{q\bar q}^2$ remains as the
relevant unknown parameter in the closed-channel approximation which
produces an uncertainty in the theoretical predictions of the
leptonic widths.  Here, in our calculations of the leptonic widths
of $\rho(nS), \rho(nD)$ with $J^{PC}=1^{--}$ we take $C_{q\bar
q}=1$.

\section{The string regime}
\label{sect.02}
In the RSH of light mesons, the quark mass $m_q=0$ and all spin-dependent
potentials are considered as a perturbation; then  the RSH is given
by the expression \cite{12,22}:
\begin{equation}
 H(\omega, \nu) = \omega + \frac{p_r^2}{\omega} +
 \frac{l(l+1)}{\omega + \int^1_0{\rm d}\beta \nu(\beta)
 \left(1 -\frac{\beta^2}{2} \right)^2} +
 \frac{\sigma^2 r^2}{2}\int^1_0{\rm d}\beta
 \frac{1}{\nu(\beta)} + \frac{1}{2} \int^1_0 {\rm d}\beta\nu(\beta).
\label{eq.4}
\end{equation}
This Hamiltonian contains two variables $\omega, \nu(\beta)$, which
are defined from the extremum conditions. The variable $\nu(\beta)$
is shown to be different for the states with large $l\geq 3$
(in the so-called the string regime) and for small $l\leq 2$ (the
potential-like regime) in order to provide the minimal value of the
mass  \cite{12,13}. In the string regime the centrifugal term  and
the term proportional to $\sigma^2 r^2$ dominate and thus the
ground state masses $M_{\rm str}(l,n_r=0)$ were obtained neglecting
the contributions from the GE and the fine structure potentials.
In that approximation the masses of all members of the multiplet
are equal to the centroid mass, which for the ground state ($n_r=0$)
with large $l$ is
\begin{equation}
 M_{\rm str}^2(l,n_r=0) = 2\pi\sigma\sqrt{l(l+1)}.
\label{eq.5}
\end{equation}
From this formula one can see that the mass difference, $\beta_l=
M_{\rm str}^2(l+1) - M_{\rm str}^2(l)$, is practically equal to
$2\pi\sigma=1.13(1)$~GeV$^2$~ ($\sigma=0.18(2)$~GeV$^2$) with high
accuracy, e.g. for $l=3$ the accuracy is $0.8\%$. The values of
$M_{\rm str}(l,n_r=0)$ are given in Table~\ref{tab.01} together
with experimental masses with $j=l+s$ and the centroid masses $M_{\rm
cog}(l,n_r=0)$ for $l=1,2$.
\begin{table}[!htb]
\caption{The masses $ M_{\rm str}(l,n_r=0)$ (in MeV) in the string regime
Eq.~(\ref{eq.4}) \label{tab.01}}
 \begin{center}
\begin{tabular}{|c|c|c|c|c|c| }
\hline
  $l$            &    1     &     2   &   3    &     4    &   5   \\
\hline
 $M_{\rm str}(l) $&  1265  &    1664    &  1979   &   2249    & 2489  \\
 $M({\rm exp.},j=l+s)  $ &   1318(1) & 1689(2) &  1982(14) &  2330(35)  & 2450(130)\\
\hline
\end{tabular}
\end{center}
\end{table}

Table~\ref{tab.01} shows the good agreement between the masses calculated
according to Eq.~(\ref{eq.5}), and the experimental masses for
$a_4(2040),~\rho_5(2350), ~a_6(2450)$ \cite{1}.  Surprisingly, even
for $a_2$ and $\rho_3$ with $l=1,2$ ,the centroid masses have reasonable
values, although the low-lying states have to be studied in the
potential-like regime and for them all kinds of the interactions:
the confining, the GE, the centrifugal term, are important. In the
potential-like regime the RSH can be rewritten in a more convenient
form, $H=H_0 + \Delta({\rm str})$, where the unperturbed part $H_0$ has
the form of the Hamiltonian occurring in the spinless Salpeter
equation (SSE) ($m_q=0$) \cite{22,23}:
\begin{equation}
 H_0 = 2\sqrt{\vep^2+ m_q^2} + V_0(r),
\label{eq.6}
\end{equation}
and the operator $p_r^2$ is replaced by $\vep^2$, while the remaining
part of the centrifugal term, the so-called string correction,
\begin{equation}
 \Delta({\rm str}) = - \frac{l(l+1)\sigma
 \langle r^{-1}\rangle_{nl}}{8 \omega^2(nl)},
\label{eq.7}
\end{equation}
is considered as a perturbation. This correction is not very large,
$\sim 50-100$~MeV for $l=1,2$, still it cannot be neglected.  In
Eq.~(\ref{eq.7}) the variable $\omega(nl)$ is the kinetic energy of a
light quark, defined by the solutions of the SSE:
\begin{equation}
(2\sqrt{\vep^2} + V_0(r))\psi_{nl}(r) =M_0(nl)\psi_{nl}(r).
\label{eq.8}
\end{equation}

To define the solutions of the unperturbed Hamiltonian $H_0$ with
$l=0,1,2$, it is important to use  the universal quark-antiquark
potential, which has no fitting parameters and therefore allows to
separate physical effects from the artifacts introduced by fitting
parameters. This potential has the form of  linear plus GE terms
(observed on the lattice \cite{33} and derived in the field correlator
method \cite{34}), and successfully describes heavy quarkonia spectra
\cite{40},
\begin{equation}
 V_0(r)= \sigma_0 r + V_{\rm GE}(r),
\label{eq.9}
\end{equation}
with $\sigma_0=0.18(2)$~GeV$^2$, fixed by the slope of the leading
RT. In the  GE potential the vector coupling in coordinate space,
\begin{equation}
 V_{\rm GE}(r)= - \frac{4}{3} \frac{\alpha_{\rm V}(r)}{r},   
label{eq.10}
\end{equation}
is taken in two-loop approximation, where it does not depend on the
renormalization scheme, and defined via the vector coupling in
momentum space:
\begin{equation}
 \alpha_{\rm V}(r)=\frac{2}{\pi} \int\limits_0^\infty{\rm d}q
 \frac{\sin(qr)}{q}\alpha_{\rm V}(q^2).
\label{eq.11}
\end{equation}
Here
\begin{equation}
 \alpha_{\rm V}(q^2) = \frac{4\pi}{\beta_0 t}
 \left(1 - \frac{\beta_1}{\beta_0^2} \frac{\ln t}{t}\right),
\label{eq.12}
\end{equation}
where for $n_f=3,~\beta_0=9,~\beta_1=64$ and in the logarithm $t(q^2)
= \ln \left(\frac{q^2 + M_{\rm B}^2}{\Lambda_{\rm V}^2}\right)$,
the vector constant $\Lambda_{\rm V}(n_f=3)=(480\pm 20)$~MeV
corresponds   to $\Lambda_{\overline{MS}}(n_f=3) = (327\pm 15)$~MeV
from pQCD, while the IR regulator $M_{\rm
B}=\sqrt{2\pi\sigma}=1.13(11)$~GeV$^2$ was defined in Ref.~\cite{38}
(see the discussion in the Introduction). At $q^2=0$ the logarithm
\begin{equation}
 t_0 = t(q^2=0) =\ln \frac{M_{\rm B}^2}{\Lambda_{\rm V}^2},
\label{eq.13}
\end{equation}
defines the freezing constant, $\alpha_{\rm crit}(q^2=0)=\alpha_{\rm
V}(r\rightarrow \infty)=0.60\pm 0.04$, which is rather large (for the
admissible values, $\Lambda_{\rm V}=480\pm 20$~MeV and $ M_{\rm
B}=1.1-1.15$~GeV ). In bottomonium, this strong GE interaction remains
important up to high excitations and  gives a good description of
the charmonium  and bottomonium spectra \cite{40}.

However, for the light mesons this universal potential appears to
be too strong, giving  smaller masses for the $1S$, $1P$, and $1D$
ground states (see Table~\ref{tab.02}). This result does not change
if the parameters of the vector coupling vary within the admissible
range. It also shows that the dynamics in light mesons, which all
lie above open hadronic thresholds, is  more complicated due to
their large spatial extensions and the appearance of virtual $q\bar
q$ loops in the Wison loop of large size, and hence modifying the
gluon exchange propagator.  As we shall discuss later in
Section~\ref{sect.05}, the gluon effectively acquires the screening
mass due to these loops as obstacles and the color-magnetic confinement
\cite{41}. For that reason we consider also a screened potential.

The creation of  the virtual quark loops (scalars in the ${}^3 P_0$
mechanism) decreases the string tension, making it dependent on the
separation $r$.  Due to this flattening effect the masses of excited
states decrease, e.g. the mass  $M(4S)$ becomes by $\sim 300-350$~MeV
smaller than for a purely linear potential  $\sigma_0 r$. However,
it is not so for the ground states with $l=0,1,2$, which have
relatively small sizes ($\langle r \rangle \leq 1.2$~fm) and are
not affected by the flattening effect.  Below, we shall find out
the direct connection between the parameter responsible for the
flattening of the potential, and the slope of the radial RT.

There is one more difference between  light meson masses and those
of heavy quarkonia, where the centroid masses just coincide with
the eigenvalue (e.v.) of the SSE. For a light meson its centroid
mass $M_{\rm cog}(nl)$ also includes a negative self-energy
contribution $\Delta({\rm SE})$ \cite{22,23,42} and negative string
correction $\Delta({\rm str})~(l=1,2)$, which do not introduce extra
parameters. The self-energy term is very important for the mass
value, since it gives contribution to the intercept of RT. In the
case  $l=0$,
\begin{eqnarray}
 M_{\rm cog}(nS) & =  & M_0(nS) +\Delta({\rm SE}),
\nonumber \\
 \Delta({\rm SE}) & = & - \frac{3\sigma}{\pi \omega(nl)}.
\label{eq.14}
\end{eqnarray}

In heavy quarkonia $\Delta({\rm SE})\sim (1-5$)~MeV is very small
and can be neglected, while for light-light,  $K$,~ and $\phi$
mesons,  $\Delta(SE)$ is rather large due to the small value of the
kinetic energy  $\omega(nl)\sim (400-500)$~MeV in the denominator.
Because of this term, the squared mass $M^2(nl)$ does not contain
a term linear in $M(nl)$)  and provides the linear behaviour of the
RT \cite{10,23}. Notice, that in the RPM a negative subtractive
constant (a fitting parameter), usually added to the potential (or
the mass) \cite{9}, violates the linearity of the orbital and radial
RT (see also the discussion in Ref.~\cite{15}).

In the mass of the $n\,{}^3S_1$ states, $M(n\,{}^3S_1)= M_0(nS) +
\Delta({\rm SE}) + \frac{1}{4} \Delta({\rm HF})$, the hyperfine
correction is defined as in Ref.~\cite{43},
\begin{equation}
 \Delta({\rm HF}) =
 \frac{32\pi \alpha_s(\mu_{\rm hf}) |\psi_{nS}(0)|^2}{9\omega^2(nS)},
\label{eq.15}
\end{equation}
where the kinetic energy $\omega(nS)$ enters in the denominator.
Here it is important to underline that in Eq.~(\ref{eq.15}) the
coupling $\alpha_s(\mu_{\rm hf})$ is not an arbitrary parameter.
As shown in Ref.~\cite{43}, this coupling is defined at the universal
scale (for all mesons, light and heavy)  $\mu_{\rm hf}\simeq
T_g^{-1}$, where $T_g\simeq 0.12$~fm is the vacuum correlation
length. Since the scale $\mu_{\rm hf}$ is close to the mass of the
$\tau$-lepton, the value of $\alpha_s(\mu_{\rm hf})$ must be close
to  $\alpha_s(M_{\tau})= 0.33(2)$ \cite{1}. We take here
$\alpha_s(\mu_{\rm hf})=0.31$, as it  was used in Ref.~\cite{44}
in the analysis of the hyperfine splitting of  the B mesons and
bottomonium.

In Table~\ref{tab.02} the calculated $\rho(nS)$ masses are  given in the typical
case, when the universal potential has no screening in the GE term
and the freezing constant $\alpha_{\rm
crit}=0.608,~~\sigma_0=0.18$~GeV$^2$.

\begin{table}[!htb]
\caption{{The masses of the $n\,{}^3S_1$ light mesons (in MeV) for the
universal potential with $\sigma=0.18$~GeV$^2$ and $\alpha_{\rm crit}=0.6086$.
Experimental data are taken from  Refs.~\cite{1,7}\label{tab.02}}}
\begin{center}
\begin{tabular}{|c|c|c|c|c|}
\hline
 $n=n_r+1 $      &    1   &    2      &   3     &   4      \\
\hline
 $M(n\,{}^3S_1)$  &   693    &   1478    &  2046   &    2510  \\
 exper. [1]  &    775    &   1465 (25)  &  1909(42)  & 2150(90)  \\
 data   [7]  &    775     &  1493(15)   &  1861 (17)  & 2254(22) \\
\hline
\end{tabular}
\end{center}
\end{table}

From Table ~\ref{tab.02} one can see (i) that the strong (universal)
GE potential gives the $\rho(775)$ mass, as well as the masses of
the $M(a_2(1318)) =1.240$~GeV, $M(\rho_3)=1.59$~GeV, smaller by
$\sim 80$~MeV than their experimental values.  (ii) On the contrary,
for high excitations, where the influence of the GE potential is
small, the masses $M(3\,{}^3S_1)$ and $M(4\,{}^3S_1)$ are by $\sim
150$~MeV and $\sim 300$  MeV larger than in experiment, irrespective
of the strength of the GE potential, if the linear $\sigma_0 r$
potential is used. Therefore one needs to look for another effect
(reason), responsible for the strong decrease of the $nS$ masses
observed in experiments.

\section{The flattening effect}
\label{sect.03}
In Section~\ref{sect.05}  we present the physical picture explaining
the flattening phenomena, observed on the lattice \cite{45} and
studied in Ref.~\cite{23}, while here we give concrete results of
our calculations with the confining potential, where the string
tension depends on the quark-antiquark separation $r$,
\begin{equation}
 \sigma(r) = \sigma_0 f(r),\quad
 \lim_{r \to \infty} \sigma(r) = \sigma_0 (1 - \gamma).
\label{eq.16}
\end{equation}
Here $\sigma_0=0.18(2)$~GeV$^2$ and the function $f(r)= 1 -
\gamma\frac{\exp(\sigma_0 (r - R_0))} {B + \exp(\sigma_0 (r -
R_0))}$, contains three parameters; two of them,  $B\cong 15-20$
and $R_0\cong (1.2-1.4)$~fm, are chosen in  such  way that the
flattening slowly starts at rather large distances, $\sim 1.2$~fm,
while the lowest lying states with  $l=0,1,2$ are not affected by
the flattening effect. At large distances, the string tension goes
to the limiting value, $\sigma_{\rm lim}=\sigma_0 (1 - \gamma)$.
Direct calculations show that the decrease of the $\rho(nS)~(n\geq
2)$ masses occur  mostly due to the flattening effect and the mass
shifts are very sensitive to the value of the parameter $\gamma$
in $\sigma(r)$. To reach agreement with experiment, the fitting
parameter $\gamma$ is to be taken in the narrow range, $\gamma=(0.43\pm
0.03)$, however, the values of $\gamma=0.40$ and 0.45 give rise to
different slopes of the $\rho$-trajectory.  If $\gamma=0.40$ is
taken, then a contribution from the screened GE potential is more
important than  for $\gamma=0.45$  (see Table~\ref{tab.03}), but
both variants have common features: 
\begin{itemize} 
\item[1. ]The linear behavior of the $\rho$-trajectory  starts with
$n_r=1$, although these RTs have slightly different slopes: for
$\gamma=0.40~(0.45)$ the slope  $\mu^2(\rho)=1.40~(1.35)$~GeV$^2$.

\item[2.] At the same time the mass difference, $\mu_1^2=M^2(\rho(2S)) -
M^2(\rho(1S))= 1.52(4)$~GeV$^2$ remains relatively large (in both
cases) and agrees  with $\mu_1^2({\rm exp.})=M^2(\rho(1465)) -
M^2(\rho(775))=(1.55\pm 0.07)$~GeV$^2$, if the central values of
the experimental mass are taken. The reason why $\mu_1^2$ is large,
is discussed below, in Section~\ref{sect.04}.

\item[3.] The choice of $\gamma$ directly determines the slope of the radial
RT and therefore it could be extracted from  the experimental masses
$M(\rho(3S))$ and $M(\rho(4S))$, if they would be  measured with better
accuracy.
\end{itemize}
In Table~\ref{tab.03} we give the masses in three cases (in all cases
$\sigma_0=0.182$~GeV$^2$, $m_q=0$): in case A there is no
screening of the GE potential, i.e., $\delta=0$; in the cases B and C the
exponential form of the screening, $V_{\rm scr}=V_{\rm GE} \exp(-\delta
r)$ with the screening parameter $\delta=0.20$~GeV, is taken.  In
the cases A and  B the other parameters coincide,
\begin{eqnarray}
 & & \Lambda_{\rm V}(n_f=3)=465~{\rm MeV},~ M_B=1.15~{\rm GeV},~
 \alpha_{\rm crit}=0.5712, ~
\nonumber \\
  & & \sigma_0=0.182~{\rm GeV}^2,~ \gamma=0.40,~ B=20,~R_0=6.0~{\rm GeV}.
\label{eq.17a}
\end{eqnarray}
In case C the stronger GE potential, with $\alpha_{\rm crit}=0.635$,
is taken, while $\gamma=0.45$ in the flattening potential is larger
than in Eq.~(\ref{eq.17a}). The other parameters in case C are as follows:
\begin{eqnarray}
 & & \Lambda_{\rm V}=500~{\rm MeV}, \quad M_B=1.15~{\rm GeV},
\nonumber \\
 & &\sigma_0=0.182~{\rm GeV}^2, \quad B= 15, \quad R_0=6.0~{\rm GeV}^2.
\label{eq.17b}
\end{eqnarray}
For all $nS$-states  the hyperfine correction to the masses is calculated with
$\alpha_s(\mu_{\rm hf})=0.31$.

As seen from Table~\ref{tab.03}, without screening ($\delta=0$) the
ground state masses of $\rho(775)$, $a_2(1318)$, and $\rho_3(1690)$
appear to be $50 - 100$ MeV  smaller than in experiment and variations
of the parameters within reasonable ranges do not change this result.
On the contrary, in the case B for the screened GE potential
($\delta=0.20$~GeV,  the other parameters remaining the same, as
in case A) a reasonable agreement with experiment is reached. The
choice of $\delta=0.30$~GeV, i.e., stronger  suppression of the GE
potential, gives rise to large  masses of  $\rho(775)$ and $\rho(1465)$
and was neglected.

The best agreement with experimental data takes place in the case C
Eq.~(\ref{eq.17b}), when the screening parameter $\delta=0.20$~GeV is the
same, but  $\Lambda_V=500$~MeV and $\gamma=0.45$ are larger than
in the cases A and B.
\begin{table}[!htb]
\caption{The masses of the lowest lying states $(l=0,1,2)$ and excited
$n\,{}^3S_1$ states (in MeV) for the flattening potential:  case
A with $\gamma=0.40,~ \delta=0$~GeV;  case B with
$\gamma=0.40,~\delta=0.20$~GeV;  case C with
$\gamma=0.45,~\delta=0.20$~GeV, $\Lambda=500$~MeV. \label{tab.03}}
\begin{center}
\begin{tabular}{|c|c|c|c|c|}
\hline
     &$\delta=0$ & $\delta=0.20$ &  $\delta=0.20$ &      \\
\hline
 state              &  Case A     &  Case B     &   Case C  &  exp. \\
 \hline
  $1\,{}^3S_1 $ &  $ 698^{a)} $   &  790    &  774  &  775   [1] \\
  $2\,{}^3S_1 $  &  1430      &   1474      & 1468 &  1465 [1] \\
               &            &         &             &   1493(15) [7]\\
  $3\,{}^3S_1 $ &  1876    &    1920    & 1880  & 1909(42) [1] \\
              &           &           &        &  1861(17)  [7] \\
 $ 4\,{}^3S_1$  &  2172    &   2239    &  2170   &  2150(90) [1] \\
              &           &          &          &   2254(22) [7] \\
  $1\,{}^3P_2  $ &  1240     &  1312    &   1309   &   1318(1) \\
 $ 1\,{}^3D_3$  & 1590     &   1696    &   1690  &    1689(2)  \\
\hline
\end{tabular}

$^{a)}$ Here the hyperfine contribution $\sim 65$~MeV is taken into account.
\end{center}
\end{table}

Our conclusion is that screening of the universal GE potential is
necessary to obtain correct values of the masses of the lowest lying
states with $l=0,1,2$, otherwise they are  $\sim 80$~ MeV smaller
than in experiment. For the higher $nS$ excitations the contribution
from the GE potential cannot be neglected and agreement with
experiment is reached both in the cases B and C. The masses of the
$nP$ and $nD~(n_r\geq 1)$ states weakly depend on the screened GE
potential  and will be discussed in the next Section.

We give here also the radii (r.m.s.) $R_s(nS)= \langle
\sqrt{r^2}\rangle_{nS}$ of $\rho(1S)$ and $\rho(2S)$, which weakly
change in all three cases: $R_s(1S)=0.71(0.72)$~fm in the cases B
(C) and a bit smaller, 0.68~fm in the case A, where there is no
screening effect. The r.m.s. of $\rho(2S)$ is significantly larger,
$R_s(2S)=1.0(1)$~fm in all cases.

\section{Radial Regge Trajectories}
\label{sect.04}
We have shown that the GE potential gives a small contribution to the
masses of the high radial excitations ($n_r\geq 1$) and therefore, in
first approximation, the GE potential with  screening can be
neglected. This allows to reveal more explicitly the role of the
flattening effect for formation of the radial RT. The most important
contribution to the light meson masses comes from the e.v. of the SSE,
Eq.~(\ref{eq.8}), the unperturbed part of the RSH,  which in the case of the
linear $\sigma_0 r$ potential ($\sigma_0$ is a  constant) is well
known. Namely, the e.v. of the SSE ($m_q=0$) can be approximated
with great accuracy (for $n_r\geq 1$) by the expression
\cite{16,23},
\begin{equation}
 M_0^2(nl)= \sigma_0 ( 8 l + 4\pi n_r  +3\pi).
\label{eq.18}
\end{equation}
This formula explicitly shows that the e.v.s $M_0^2(n_r=0,l)$ for
the ground states for given $l$, lie on the orbital  RT with the
slope $\beta_0=8\sigma_0=1.44$~GeV$^2$, which is $\approx 27\%$
larger than $\beta({\rm exp.})= 2\pi\sigma_0=1.13(1)$~GeV$^2$,
observed in experiment.  For the radial excitations, the difference
between the slope in Eq.~(\ref{eq.18}) and the one found  in
experiment is very large: $\mu_0^2= 4\pi\sigma_0 = 2.26$~GeV$^2$
is $1.6-2.0$ times larger than $\mu^2({\rm exp.})=(1.25\pm
0.15)$~GeV$^2$ \cite{8,11}. The question is why such a large
difference occurs.

First of all, we look at the contribution to the centroid mass from
the self-energy correction, Eq.~(\ref{eq.14}), for which we use the
relation, $M_0(nS) = 4\omega_0(nS)$ (valid for $\sigma=\rm const.$)
and rewrite $\Delta({\rm SE})= -\frac{12\sigma_0}{\pi M_0}$; then
\begin{equation}
 M_{\rm cog}(nS) = M_0(nS)- \frac{3.82\sigma_0}{ M_0(nS)}.
\label{eq.19}
\end{equation}
In the squared mass we neglect the small squared self-energy term
(although it is not small for the $1S$ state)  and obtain
\begin{equation}
 M_{\rm cog}^2(nS) = M_0^2(nS) - 7.6 \sigma_0 =
 \sigma_0 (4\pi n_r - 7.6 +3\pi)= (0.33 +2.26 n_r)~{\rm GeV}^2.
\label{eq.20}
\end{equation}
From here one can see that owing to $\Delta(SE)$ the value of $M_{\rm
cog}^2$ is smaller than $M_0^2$ given in Eq.~(\ref{eq.18}), while
the slope $\mu_0^2=4\pi\sigma_0=2.26(2)$~GeV$^2$ does not change.
Thus we have confirmed the well-known result that the purely linear
potential with $\sigma=\rm const.$ produces always a large slope
of the radial RT.

The situation strongly changes, if the flattening potential $V_{\rm
C}(r)= \sigma(r) r$ is considered, for which the representation
Eq.~(\ref{eq.18}) is not valid anymore (in this case the e.v. of the SSE will
be denoted as $\tilde{M}_0(nS)$). Our calculations show that
\begin{itemize}
\item[1.] The linear behavior of the $\rho$ RT starts with $n_r=1$, because
for the flattening potential (with $\gamma=0.40$ or $0.45$) the mass
difference $\mu_1^2 = \tilde{M}^2(2S) - \tilde{M}^2(1S)$ remains
large, $\mu_1^2\sim 1.87(5)$~GeV$^2$, being still  20\% smaller
than $\mu_1^2=4\pi\sigma_0$ in Eq.~(\ref{eq.18}).

\item[2.] For the $nP$ and $nD$ states the linear behaviour starts with $n_r=0$.

\item[3.] The slope $\mu^2(l)$ strongly depends on the parameter $\gamma$
in $\sigma(r)$ , Eq.~(\ref{eq.16}), which characterizes the weakening of the
confining potential.
\end{itemize}

The squared e.v., $\tilde{M}_0^2(nS)$ (in GeV$^2$) $(n_r\geq 1$),
with $\gamma=0.40, 0.45, 0.50$ can be approximated as
\begin{eqnarray}
 \tilde{M}_0^2 & = & (2.42 +1.40\,n_r)~{\rm GeV}^2,~{\rm for}~\gamma=0.40,
\nonumber \\
\tilde{M}_0^2  & =  & (2.31 + 1.27\, n_r)~{\rm GeV}^2,~{\rm for}~\gamma=0.45,
\nonumber \\
\tilde{M}_0^2  & =  & (2.25 + 1.15\, n_r)~{\rm GeV}^2,~{\rm for}~ \gamma=0.50.
\label{eq.21a}
\end{eqnarray}
The accuracy of these expressions is $\sim 1\%$.

From Eq.~(\ref{eq.21a}) the important result follows that for the
flattening potential the squared e.v.s of the SSE have a much smaller
slope (two times smaller for $\gamma=0.50$), than in the case of
the purely linear potential Eq.~(\ref{eq.18}), which decreases for
larger values of $\gamma$, i.e., a stronger flattening effect. For
the centroid mass the intercept is changed, while the value of the
slope is the same, so that the $\rho$ trajectory ($n_r\geq 1$) is,
\begin{eqnarray}
 M_{\rm cog}^2(\rho) & = & (0.77 +1.40\, n_r)~{\rm GeV}^2,~{\rm for}~
\gamma=0.40,
\nonumber \\
 M_{\rm cog}^2 & = & (0.80+ 1.27\, n_r)~{\rm GeV}^2,~{\rm for}~\gamma=0.45,
\nonumber \\
 M_{\rm cog}^2 & = & (0.90 + 1.15\, n_r)~{\rm GeV}^2,~{\rm for}~\gamma=0.50.
\label{eq.21b}
\end{eqnarray}
In all cases $M_{\rm cog}(\rho(1450))=(1.44-1.47)$~GeV. However, if the
exponential screening of $V_{\rm GE}$ and the hyperfine interaction are
taken into account, then the slope increases while the intercept does
practically not change. In the cases A and C (see the parameters of the
GE potential in Eqs.~(\ref{eq.17a},\ref{eq.17b})) and  for $n_r \geq 1$
we have
\begin{eqnarray}
 M^2(n\,{}^3S_1) & = & (0.78 + 1.40(2)\, n_r)~{\rm GeV}^2~ (\gamma=0.40),
\nonumber \\
 M^2(n\,{}^3S_1) & = & (0.81 + 1.34(1)\, n_r)~{\rm GeV}^2~ (\gamma=0.45).
\label{eq.22}
\end{eqnarray}
Thus for $\gamma=0.45$ the calculated $\rho$-trajectory has $\mu^2=
1.34(1)$~GeV$^2$, in agreement with the results in Refs.~\cite{11,26},
where $\mu^2=1.365(108)$~GeV$^2$ was obtained from the analysis of
the Crystal Barrel data \cite{2}. On the contrary, the larger
$\mu^2=1.40(2)$~GeV$^2$ for $\gamma=0.40$ agrees with  the slope,
$\mu^2=1.43(13)$~GeV$^2$, predicted in Ref.~ \cite{27}. Notice, that for
$\gamma=0.45$ a better agreement is obtained for the $\rho(1450)$
mass (see Table~\ref{tab.03}).

In the same way, the radial RTs for the $nP$ and $nD$ states were
considered; it appears that for $l=1,2$ the linear behavior of the
radial RT starts with $n_r=0$ and the squared e.v. of SSE $M_0^2(nP)$
can be approximated as
\begin{equation}
 \tilde{M}_0^2(nP)= (3.11 + 1.25\, n_r)~{\rm GeV}^2~ (\gamma=0.45).
\label{eq.23a}
\end{equation}
Then, taking into account the self-energy and string corrections we
obtain for the centroid masses,
\begin{equation}
 M_{\rm cog}^2(nP)=(1.64(2) + 1.25\, n_r)~{\rm GeV}^2.
\label{eq.23b}
\end{equation}
From this expression one can obtain the $a_{\rm j}$ radial RT,
taking into account the fine-structure splitting, which does
practically not change the slope, but introduces a fitting parameter. For
that reason we restrict ourselves to the  RTs for the centroid masses.
Notice that $\mu^2(nP)=1.25$~GeV$^2$ practically coincides with
the slope for the centroid masses of the $nS$ states, if $\gamma=0.45$.

For the $nD$ trajectory  the e.v.s of the SSE have a smaller slope
($n_r\geq 0$),
\begin{equation}
 \tilde{M}_0^2(nD)=(4.36 + 1.11(5)\, n_r)~{\rm GeV}^2 ~(\gamma=0.45).
\label{eq.24a}
\end{equation}
and
\begin{equation}
 M_{\rm cog}^2(nD) = (2.8(1) + 1.11(5)\, n_r)~{\rm GeV}^2 (\gamma=0.45).
\label{eq.24b}
\end{equation}
Notice, that the slope $\mu^2(l)$ decreases for increasing
angular-momentum $l$.  At this point it is important to stress that
for physical $nD$ states the GE contribution is much smaller than
that for the $nS$ states, and therefore for the $\rho_3$, $\rho_2$,
and $\rho(n\,{}^3D_1)$ trajectories the slopes have to be close to
the one given in  Eq.~(\ref{eq.24b}), where $\mu^2(D)=1.11(5)$~GeV$^2$,
if the fine-structure effects are neglected. Our result is in
agreement with  $\mu^2(a_2)=1.00(6)$~GeV$^2$ and $\mu^2(a_1)=(1.084\pm
0.63)$~GeV$^2$, predicted for the $a_1$ and $a_2$ RTs  from the
analysis of experimental data in Ref.~\cite{26}.

Now we briefly discuss the reasons why the mass difference
$\mu_1^2=M^2(2S)- M^2(1S)$, is large. The first reason is that this
factor is very large for the flattening potential (without GE
interaction), where  $\mu_1^2=1.87(5)$~GeV$^2$ for the squared e.v.
of SSE. This result does not change, if a reasonable choice of the
parameters in $\sigma(r)$ is made.  Secondly, if the GE interaction
is taken into account, then for the $1S$ state, localized at rather
small distances, the self-energy and hyperfine corrections decrease
the mass difference $\mu_1^2$ but its value remains  rather large,
$\mu_1^2=1.56(6)$~GeV$^2$.  This number appears to be very close
to what is observed in experiment, $\mu_1^2(\rm exp.)=1.55(7)$~GeV$^2$
\cite{1}, if the central values of the $\rho(775)$ and $\rho(1465)$
masses are used. Just for that reason, the linear behaviour of the
$\rho$ RT begins with $n_r=1$, while other radial RTs start with
$n_r=0$.

\section{Flattening phenomenon - the physical picture}
\label{sect.05}
The dynamics of light mesons is more complicated than that in heavy
quarkonia, since light mesons, as rather extended objects, are sensitive
to detailed properties of the confinement mechanism, which also
affects the gluon exchanges. Our approach is based  on the background
perturbation theory (BPT) \cite{46}, which takes into account the
non-perturbative background with confinement and does not contain
unphysical singularities (the Landau ghost poles and IR renormalons),
present in standard perturbation theory. Below, we illustrate
how the BPT predicts three effects, which are observed in experiment
and especially important for light mesons:
\begin{itemize}
\item[1.] Stabilisation of the coupling $\alpha_s(q^2)$ at
$q^2\rightarrow 0,~\alpha(0)\equiv \alpha_{\rm crit}$;

\item[2.] Screening of $\alpha_s$ at large distances;

\item[3.] Flattening of the string tension at large distances,
$\sigma\rightarrow \sigma(r)$.  
\end{itemize} 
Item 1. The basic feature of BPT is the gauge-invariant treatment
of confinement and gluon-exchanges, when both phenomena occur owing
to the Wilson loop, where confinement creates the minimal-area
surface (the so-called confining film) and the gluon-exchange
trajectories are necessarily present inside this surface. As a
result, the gluon loops, appearing on these trajectories and
responsible for asymptotic freedom, create open loops in the confining
film, and this effect strengthens with increasing $\alpha_s$, leading
finally to the saturation of $\alpha_s(q^2)$ which in two-loop
approximation is given by $\alpha_{\rm crit}=\frac{4\pi}{\beta_0
t_0} \left( 1 - \frac{\beta_1}{\beta_0^2} \frac{\ln t_0}{t_0}\right)$,
where $t_0=\ln(\frac{M_{\rm B}^2}{\Lambda^2})$ and with $10\%$
accuracy $M_{\rm B}^2=2\pi\sigma$ \cite{41}.

Item 2. For the same reason, the scalar $q\bar q$ loops, appearing
in the film (of large size), lead to the screening of the GE
interaction, since any gluon trajectory, propagating inside the
confining film, is interrupted by the scalar loops and those create
an effective mass of the gluon. In addition, there is a difference
between the free propagation (free Green's function of the gluon)
and the gluon propagation inside the surface with confinement. This
complicated phenomenon was studied in Ref.~\cite{47} for zero
temperature and in Ref.~\cite{48} for the deconfined phase, where
it occurs due to the color-magnetic confinement.  This effect at
zero temperature, when both color-electric and color-magnetic
confinement collaborate, is not yet finally settled and therefore
in our paper we exploit the effective screening parameter $\delta$
for the screening mass. Our analysis has shown that for the exponential
form of screening $\delta=0.20$~GeV is the preferable value, while
suppression of the GE potential is too strong for $\delta=0.30
(0.10)$~GeV.

Item 3. For excited light mesons, confinement occurs in a highly
excited string, when the Wilson loop has a free boundary and several
typical features, partly discussed above. Namely, there exist

(i)  a finite density of the $q\bar q$ loops in the confining film,
which leads to the dependence of the string tension on $r$,
$\sigma\rightarrow \sigma(r)$;

(ii)  a possibility to decay, virtually or really, into a pair (or
several) mesons, so that if the distance $r$ in the confining
potential $\sigma r$ exceeds the separation, $R_{\rm f}\sim 2
r_{\pi}\simeq 1.2$~fm, then the flattening of the potential is
expected.

The first feature can also be seen in the  $T$-dependence of
$\sigma(r)$: when the density of the $q\bar q$ loops grows with
increasing temperature $T$,  then the potential $V(r,T)$ becomes
more and more flat, as it was observed on the lattice \cite{49}.
Another manifestation of the flattening phenomenon was recently
studied in Ref.~\cite{50}:  while applying a magnetic field parallel
to the confining film, it was observed that the density of the
$q\bar q$ loops increases and the string tension $\sigma(r)$ flattens,
in agreement with the lattice data \cite{51}.

Both features, flattening due to a finite $q\bar q$ density and the
existence of the critical length $R_{\rm f}\sim 2 r_{\pi}$ are
embodied in the form of the string tension, Eq.~(\ref{eq.16}), which
is used in our paper.

In conclusion we give  the r.m.s of the $\rho(nS)$ mesons, $r_s=\langle
\sqrt{r^2}\rangle_{nS}$, calculated for the sets of the parameters
Eqs.~(\ref{eq.17a}, \ref{eq.17b}): for $\rho(775)$,
$r_s(1S)=(0.71-0.73)$~fm and for  $\rho(1450)$, $r_s=(0.9-1.0)$~fm.

\section{The leptonic widths of $\rho(n\,{}^3S_1)~l=0,2$}
\label{sect.06}
The decay constants $f_{\rm V}$ and leptonic widths of the
$\rho(n\,{}^3S_1)$ mesons are calculated here,  considering them
as purely $q\bar q$ states, i.e., taking  $C_{q\bar q}=1.0$ in the
w.f. given in Eq.~(\ref{eq.3}). For the decay constant in the vector
channel $f_{\rm V}$ we use the expression from Ref.~\cite{52}, where
the correlator of the currents (in different channels) is derived
using the functional integral representation and on the final stage
expanding this correlator in the complete set of  eigenfunctions
of the RSH $H_0$, Eq.~(\ref{eq.6}). This  gives
\begin{equation}
 f_{\rm V}^2 = 12\bar{e}_q^2 \frac{|\psi_n(0)|^2 \xi_{\rm V}}{M_{\rm V}(nS)}=
 \frac{3 \bar{e}_q^2|R_n(0)|^2 \xi_{\rm V}}{\pi M_{\rm V}(nS)},
\label{eq.25}
\end{equation}
and
\begin{equation}
 \Gamma_{\rm ee}(n\,{}^3S_1) =
 \frac{4\pi\alpha^2 f_{\rm V}^2 \beta_{\rm QCD}}{3 M_{\rm V}}.
\label{eq.26}
\end{equation}
Here, for a light meson with $m_q=0$ the relativistic factor
$\xi_{\rm V}(nS)$  is
\begin{equation}
 \xi_{\rm V} = \frac{\omega_n^2 + \frac{1}{3} \vep^2}{2 \omega_n^2},
\label{eq.27}
\end{equation}
which for the ground and excited states are almost equal,
$\xi(1S)=0.70(1)$ and $\xi(nS)=0.72(1),~ (n=2,3,4)$, if the static
potential with the parameters Eqs.~(\ref{eq.17a},\ref{eq.17b}) is
used (for the $\rho$-mesons the average $\bar{e}_q^2 =1/2$).  The
factor $\beta_{\rm QCD}=1-\frac{16}{3\pi} \alpha_s =0.40$ takes
into account the radiative corrections \cite{53} and here we use
for all ($n\,{}^3S_1$)-states the same coupling $\alpha_s(\mu_s)=0.353$
(at the scale  $\mu_s\sim 1.0$~GeV).  If the confining potential
flattens at large distances, then the w.f.s at the origin $R_{\rm
nS}(0)\sim (0.36-0.33)$~GeV$^{3/2}~(n=2-4)$ have close values ,
while for the ground state  the w.f. $R_{\rm 1S}(0)=(0.376\pm
0.008)$~GeV$^{3/2}$ is larger, and for the $\rho(775)$  the decay
constant and leptonic width are
\begin{equation}
 f_{\rm V}=(245\pm 6)~{\rm MeV},
\label{eq.28a}
\end{equation}
where the uncertainty comes from that in the w.f. at the origin, and
\begin{equation}
 \Gamma_{\rm ee}(\rho(775))=( 7.0\pm 0.3)~{\rm keV}.
\label{eq.28b}
\end{equation}

To calculate the leptonic widths of the higher $\rho(nS)$, it is
convenient to use the ratio of the leptonic widths,
$\Gamma_{\rm ee}(n\,{}^3S_1)/\Gamma_{\rm ee}(\rho(775))$, where the
factors $\xi(nS)$ and $\beta_{\rm QCD}$ drop out. This gives
\begin{eqnarray}
 \Gamma_{\rm ee}(2\,{}^3S_1) & = & 0.24\;\;
 \Gamma_{\rm ee}(\rho(775))= 1.7(1)~{\rm keV},
\nonumber \\
 \Gamma_{\rm ee}(3\,{}^3S_1) & = & 0.14\;\;
 \Gamma_{\rm ee}(\rho(775))=1.0(1)~{\rm keV},
\nonumber \\
 \Gamma_{\rm ee}(4\,{}^3S_1) & = & 0.096\;
 \Gamma_{\rm ee}(\rho(775))=0.7(1)~{\rm keV},
\label{eq.29}
\end{eqnarray}
where the uncertainties come from the experimental errors in the
$\rho(nS)$ masses and the w.f.s at the origin. Notice that in a
realistic situation the leptonic widths of the excited $\rho(nS)$
mesons may be smaller, if the $q\bar q$ component $C_{q\bar q}$ in
their w.f.s is less than 1.0.

The leptonic widths of the ($n\,{}^3D_1$)-states is calculated
defining their w.f.s at the origin via the second derivative, according
to the prescription  from Ref.~\cite{54}: $R_{nD}(0)=\frac{5
R''(0)}{2\sqrt{2}\omega_{nD}^2}$, where $R_{1D}''(0)=0.026(1)$~GeV$^{7/2}$
and $\omega(1D)=0.536$~GeV. It gives $R_{1D}(0)=0.163(3)$~GeV$^{3/2}$,
which is not very small due to the small value of the kinetic energy
$\omega_{nD}\sim 0.5$~GeV. The other parameters are $\xi(1D)=0.69,~
\beta_{\rm QCD}=0.40,~ M(1\,{}^3D_1)=1.72(2)$~GeV, so that the
leptonic width,
\begin{equation}
 \Gamma_{\rm ee}(1\,{}^3D_1)=0.26(5)~{\rm keV}
\label{eq.30}
\end{equation}
is rather small. However, its value may increase owing to the $2S-1D$
mixing, and for a mixing angle $\theta=21^\circ$ the leptonic widths
of  $\rho(1450)$ and $\rho(1700)$ become almost equal:
\begin{equation}
 \Gamma_{\rm ee}(\rho(1450))=1.0(1)~{\rm keV},\quad
 \Gamma_{\rm ee}(\rho(1700))=0.75(6)~{\rm keV}~(\theta=21^\circ).
\label{eq.31}
\end{equation}
Here it was assumed that in the w.f.s of $\rho(1450)$ and $\rho(1700)$
the $q\bar q$ components are equal.

\section{Conclusions}
\label{sect.07}
We have studied  the light meson properties with the use of the
RSH, which allows us to investigate the light-meson dynamics without
introducing fitting parameters.  It appears that the universal
static potential, successfully applied to heavy quarkonia, gives
rise to small masses of the lowest states with $l=0,1,2$ and at the
same time large masses of the excited states. To explain the physical
spectrum, two effects: the screening of the GE interaction and the
flattening of the confining potential, which appear owing to
quark-loop creation, are to be taken into account. We have demonstrated
the following properties.

\begin{itemize}
\item[1.] The screening of the GE potential, taken as an exponential
function  with the screening parameter $\delta=0.20$~GeV, gives the
masses of the lowest lying states for each $l$ in agreement with
experiment.

\item[2.] The slope of the radial RT is very sensitive to the value
of the parameter $\gamma$, which determines the flattening of the
string tension $\sigma(r)$:  at large distances  $\sigma(r)\rightarrow
\sigma_0(1 - \gamma)$. The parameter $\gamma$ could be extracted
from the experimental masses of $\rho(1900)$ and $\rho(2150)$, if
these were measured with better accuracy, while now it is taken
from the range $\gamma=0.43\pm 0.03$.

\item[3.] From our calculations two values for the  slope of the
$\rho$-trajectory are obtained, $\mu^2(\rho)=1.40(2)$~GeV$^2$ for
$\gamma=0.40$ and $\mu^2(\rho)=1.34(1)$~GeV$^2$ for $\gamma=0.45$,
neither result contradicts the existing experimental data.

\item[4.] The linear behaviour of the radial RT starts with $n_r=0$
for the $nP$ and $nD$ trajectories, while the linear behaviour of
the $\rho$ trajectory begins with the first excitation, $n_r=1$,
since the large value of the mass difference,
$M^2(\rho(1450))-M^2(\rho(775))=1.56(6)$GeV$^2$ (or the relatively
small value of the $\rho(775)$ mass) is a dynamical property of the
$1S$ ground state.

\item[5.] The leptonic widths $\Gamma_{\rm ee}(\rho(775))=7.0(3)$~keV,
$\Gamma_{\rm ee}(\rho(1450))=1.7(1)$~keV, $\Gamma_{\rm
ee}(\rho(1900))=1.0(1)$~keV, $\Gamma_{\rm ee}(2150))=0.7(1)$~keV,
and $\Gamma_{\rm ee}(\rho(1700)) =0.26(5)$~keV are calculated
considering them as purely $q\bar q$ states. If $2S-1D$ mixing is
possible, then for the mixing angle $\theta=21^\circ$ comparable
values of the leptonic widths $\Gamma_{\rm ee}(\rho(1450))=1.0(1)$~keV
and $\Gamma(\rho(1700)=0.75(6)$ ~keV are obtained.

\end{itemize}

\begin{acknowledgments}
The financial support of the grant RFBR 1402-00395 is gratefully
acknowledged by A.~M.~Badalian.
\end{acknowledgments}

\end{document}